  \documentstyle[twocolumn,pre,aps,epsf]{revtex}

\def\vector#1{{\bf #1}}

\def\vk{{\vector k}}
\def\vq{{\vector q}}

\def\vR{{\vector R}}

\def\va{{\vector a}}
\def\vb{{\vector b}}

\def\dps{\displaystyle}

\def\Tc{{T_{\rm c}}}

\def\hightc{{high-$T_{\rm c}$ }}

\def\Tc{{T_{\rm c}}}

\def\TMTSFX{{${\rm (TMTSF)_2X}$}}
\def\TMTSFPF{{${\rm (TMTSF)_2PF_6}$}}
\def\TMTSFClO{{${\rm (TMTSF)_2ClO_4}$}}

\def\SrRuO{{${\rm Sr_2RuO_4}$}}

\begin{document}
\draft

\twocolumn[\hsize\textwidth\columnwidth\hsize\csname 
@twocolumnfalse\endcsname


\title{Coexistence of Singlet and Triplet Attractive Channels 
       in the Pairing Interactions Mediated 
       by Antiferromagnetic Fluctuations}

\author{Hiroshi Shimahara} 

\def\runtitle{}


\address{
Department of Quantum Matter Science, ADSM, Hiroshima University, 
Higashi-Hiroshima 739-8526, Japan
}

\date{Received ~~~ March 2000}

\maketitle

\begin{abstract}
We propose a phase diagram of quasi-low-dimensional type II superconductors 
in parallel magnetic fields, when antiferromagnetic fluctuations 
contribute to the pairing interactions. 
We point out that pairing interactions mediated by antiferromagnetic 
fluctuations necessarily include both singlet channels and triplet channels 
as attractive interactions. 
Usually, a singlet pairing is favored at zero field, 
but a triplet pairing occurs at high fields where the singlet pairing is 
suppressed by the Pauli paramagnetic pair-breaking effect. 
As a result, the critical field increases divergently at low temperatures. 
A possible relation to experimental phase diagrams of a quasi-one-dimensional 
organic superconductor is briefly discussed. 
We also discuss a possibility that a triplet superconductivity is observed 
even at zero field. 
\end{abstract}



]

\narrowtext

Pairing interactions mediated by a spin fluctuations have been 
discussed as a possible mechanism of anisotropic 
superconductivity~\cite{Eme86,Bea86,Sca86,Miy86,Miy88,Shi88,Shi89,Yon90,Ued92} 
in exotic superconductors such as organic superconductors, 
heavy fermion superconductors, \hightc cuprate superconductors, 
and ruthenate superconductors. 
For example, in a ruthenate superconductor \SrRuO, 
pairing interaction mediated by ferromagnetic fluctuations 
is a candidate for the mechanism of the superconductivity. 
The superconductivity in this compound is considered to be due to triplet 
pairing from experimental results such as those obtained in Knight shift 
measurements~\cite{Ish98} and $\mu SR$ experiments~\cite{Luk98}.

On the other hand, in the organics, cuprates, and heavy fermion 
superconductors, 
anisotropic singlet pairing intereactions mediated by antiferromagnetic 
fluctuations have been examined by many authors, 
because these compounds are in proximities to antiferromagnetic phases. 
However, in some of the quasi-one-dimensional organic superconductors 
and the heavy fermion superconductors, 
triplet superconductivities have been supported by 
some experimental results~\cite{Bel97,Lee00}. 
In particular, in some of the heavy fermion superconductors, 
the triplet pairing seems to be established.

In a quasi-one-dimensional organic superconductor \TMTSFPF, 
triplet pairing is supported by recent Knight shift measurements 
by Lee {\it et al.}~\cite{Lee00}, but it may appear to contradict 
the proximity to an antiferromagnetic phase. 
We will briefly discuss later that it does not necessarily contradict 
the mechanism of superconductivity with pairing interactions mediated 
by antiferromagnetic fluctuations. 
There is another contradiction between the thermal 
conductivity measurements~\cite{Bel97} and the NMR 
experiments~\cite{Tak87,Has87} in \TMTSFClO. 
The former indicates a nodeless gap, 
while the latter suggests an existence of line nodes. 
However, this contradiction can be consistently explained also in terms 
of the pairing interactions mediated by antiferromagnetic 
fluctuations~\cite{Shi00}.

In the cuprate \hightc superconductors, experiments with $\pi$-junction 
suggest an anisotropic pairing which is conventionally 
called a $d$-wave pairing. 
Pairing interactions mediated by spin fluctuations have been discussed 
by many authors since the discovery of the \hightc superconductors, 
for example, Miyake {\it et al.}~\cite{Miy88} 
and the present author {\it et al.}~\cite{Shi88}. 
However, it is not established to what extent the spin fluctuations 
contribute to the pairing interactions.

In this paper, we discuss a phase diagram 
of quasi-low-dimensional type II superconductors 
on a $T$-$H$ plane. 
We consider a situation in which a singlet pairing is favored 
at zero field because of pairing interactions induced 
by the exchange of antiferromagnetic spin fluctuations. 
We restrict our discussion to parallel magnetic fields so that 
we can regard the orbital pair-breaking effect as weak.

The pairing interactions mediated by spin fluctuations 
are written as 
\def\eqPISF{(1)}
$$
     H' = - g \sum_{ij}
     \chi_{ij} {\vec S}_i \cdot {\vec S}_j  , 
     \eqno\eqPISF
     $$
where $i$ and $j$ denote lattice sites 
and $\chi_{ij}$ is the expression of the spin susceptibility 
in real space. 
The dynamical effects and strong coupling effects can be taken into 
account by developing a perturbation theory in more basic models 
such as Hubbard models~\cite{Shi88,Shi89,Yon90}. 
However, since it is not essential for the purpose of this paper, 
we discuss pairing interactions of the form of eq.~{\eqPISF} 
for a while. Equation {\eqPISF} can be rewritten as 
\def\eqPISFst{(2)}
$$
\renewcommand{\arraystretch}{2.0}
     \begin{array}{rcl}
     H' 
     & = & \dps{ 
     - \frac{g}{N} \sum_{\vk \vk'}
     \chi(\vk-\vk') 
     {\Bigl [} 
     \frac{1}{4} 
     {\bigl (} 
     \psi_{11}^{\dagger}(\vk)
     \psi_{11}(\vk')
     } \\
     && \dps{ 
     + 
     \psi_{1,-1}^{\dagger}(\vk)
     \psi_{1,-1}(\vk')
     + 
     \psi_{10}^{\dagger}(\vk) \psi_{10}(\vk')
     {\bigr )}
     } \\
     && \dps{ 
     - 
     \frac{3}{4}
     \psi_{00}^{\dagger}(\vk) \psi_{00}(\vk')
     {\Bigr ]} 
     } 
     \end{array}
     \eqno\eqPISFst
     $$
where 
\def\eqpsidef{(3)}
$$
\renewcommand{\arraystretch}{1.5}
     \begin{array}{rcl} 
     \psi_{11}(\vk)   & = & c_{-\vk \uparrow} c_{\vk \uparrow}       \\
     \psi_{1,-1}(\vk) & = & c_{-\vk \downarrow} c_{\vk \downarrow}   \\
     \psi_{10}(\vk)   & = & \frac{1}{\sqrt{2}} 
     (  c_{-\vk \uparrow} c_{\vk \downarrow}
      + c_{-\vk \downarrow} c_{\vk \uparrow} ) \\
     \psi_{00}(\vk)   & = & \frac{1}{\sqrt{2}} 
     (  c_{-\vk \uparrow} c_{\vk \downarrow}
      - c_{-\vk \downarrow} c_{\vk \uparrow} )   .  \\
     \end{array}
     \eqno\eqpsidef
     $$

When the antiferromagnetic fluctuations are strong, the factor $\chi_{ij}$ 
changes its sign depending on the distance between the sites $i$ and $j$. 
In general, the factor $\chi_{ij}$ is positive for the sites $(i,j)$ 
on the same sublattice, while $\chi_{ij}$ is negative 
for the sites $(i,j)$ on the different sublattices, 
for the antiferromagnetic correlations. 
Here, it is easily verified from eq.~{\eqPISFst} 
that the spin correlations between the different sublattices 
($\chi_{ij} < 0$) favor singlet pairing, 
while they on the same sublattices ($\chi_{ij}>0$) favor triplet pairing.

For example, in tight binding models near the half-filling, 
the factor $\chi_{ij}$ changes its sign alternately. 
For the nearest-neighbor sites $(i,j)$, $\chi_{ij} \equiv {\bar \chi}_1 < 0$, 
while for the next-nearest-neighbor sites $\chi_{ij} \equiv {\bar \chi}_2 > 0$. 
The former ${\bar \chi}_1 < 0$ favor the singlet pairing, 
while the latter ${\bar \chi}_2 > 0$ contribute to 
the next-nearest-neighbor triplet pairing.

The triplet pairing interactions between the next-nearest-neighbor sites 
have not been considered to be important so far, 
because the nearest-neighbor singlet pairing interactions are usually 
much stronger than the next-nearest-neighbor triplet pairing interactions. 
However, we could not ignore the triplet pairing interactions between 
electrons with parallel spins at high fields where the singlet pairing 
is suppressed by the Pauli pair-breaking effect.

It might be considered that the contributions to the triplet pairing 
interactions from ${\bar \chi}_2 > 0$ may be completely canceled 
by the larger negative contributions from ${\bar \chi}_1 < 0$. 
In order to illustrate that an attractive channel remains as a total, 
let us consider an example in which there are only two kinds of $\chi_{ij}$ 
with opposite signs, 
and write them ${\bar \chi}_1$ and ${\bar \chi}_2$. 
For example, in a one-dimensional tight binding model, 
if we truncate $\chi_{ij}$ at the next-nearest-neighbor sites, 
we obtain ${\bar \chi}_1 < 0$ and 
${\bar \chi}_2 > 0$ and $|{\bar \chi}_1| > |{\bar \chi}_2|$.

Then, the gap equations are written in a matrix form 
\def\eqgapeq{(4)}
$$
     {\left (
     \begin{array}{c}
     \Delta_1 \\
     \Delta_2 
     \end{array}
     \right )}
     = 
     - V 
     {\left (
     \begin{array}{cc}
     {\bar \chi}_1 W_{11} & {\bar \chi}_1 W_{12} \\
     {\bar \chi}_2 W_{21} & {\bar \chi}_2 W_{22}
     \end{array}
     \right )}
     {\left (
     \begin{array}{c}
     \Delta_1 \\
     \Delta_2 
     \end{array}
     \right )} , 
     \eqno\eqgapeq
     $$
where $V = 3g/8 > 0$ for the singlet pairing 
and $V = - g/8 < 0$ for the triplet pairing. 
The matrix elements $W_{nm}$ are defined by 
\def\eqWnmdef{(5)}
$$
     W_{nm} = \frac{1}{N} \sum_{\vk} 
     \gamma_n(\vk) \frac{\tanh\frac{\epsilon_{\vk}}{2T}}{2\epsilon_{\vk}}
     \gamma_m(\vk)  , 
     \eqno\eqWnmdef
     $$
where $\gamma_n(\vk)$ are the momentum dependent factors to expand 
the gap function. 
For example, when we consider one dimensional systems with interactions 
up to the next-nearest-neighbor sites, $\gamma_n(\vk)$ are defined by 
\def\eqgammadefsinglet{(6)}
$$
     \begin{array}{rcl}
     \gamma_1(k_x) & = & \sqrt{2} \cos k_x   \\
     \gamma_2(k_x) & = & \sqrt{2} \cos 2 k_x
     \end{array} 
     \eqno\eqgammadefsinglet
     $$
for the singlet pairing, and 
\def\eqgammadeftriplet{(7)}
$$
     \begin{array}{rcl}
     \gamma_1(k_x) & = & \sqrt{2} \sin k_x   \\
     \gamma_2(k_x) & = & \sqrt{2} \sin 2 k_x
     \end{array} 
     \eqno\eqgammadeftriplet
     $$
for the triplet pairing. 
Extensions to two dimensional systems are straightforward. 
For example, $\gamma_n$ are defined by 
\def\eqgammatwoddefdwave{(8)}
$$
     \begin{array}{rcl}
     \gamma_1(\vk) & = & \cos k_x - \cos k_y     \\
     \gamma_2(\vk) & = & \cos 2 k_x - \cos 2 k_y \\
     \end{array} 
     \eqno\eqgammatwoddefdwave
     $$
for a $d$-wave-like pairing, while they are defined by 
\def\eqgammatwoddefpwave{(9)}
$$
     \begin{array}{rcl}
     \gamma_1(\vk) & = & \sqrt{2} \sin k_x \\
     \gamma_2(\vk) & = & 2 \sin k_x \cos k_y \\
     \end{array} 
     \eqno\eqgammatwoddefpwave
     $$
for a $p$-wave-like pairing. 
The matrix elements $W_{12}=W_{21}$ are not equal to zero 
for non-half-filling. 

Now, we show that the two eigen values of the matrix on the right hand side 
of eq.~{\eqgapeq} have opposite signs in general. 
They are equal to 
\def\eqeigenvalue{(10)}
$$
\renewcommand{\arraystretch}{2}
     \begin{array}{rcl}
     \lefteqn{ \dps{ 
     \lambda_{\pm} 
     = 
     \frac{1}{2}
     {\bigl [}
     (  {\bar \chi}_1 W_{11} + {\bar \chi}_2 W_{22} ) 
     } } \\
     && \dps{ 
     \pm 
     \sqrt{
     (  {\bar \chi}_1 W_{11} + {\bar \chi}_2 W_{22} )^2 
     - 4 {\bar \chi}_1 {\bar \chi}_2 
     (W_{11}W_{22} - W_{12}^2) }
     {\bigr ]} . 
     }\\
     \end{array}
     \eqno\eqeigenvalue
     $$
Since ${\bar \chi}_1 {\bar \chi}_2 < 0$ and $W_{11}W_{22} - W_{12}^2 > 0$, 
the second term in the square root is positive 
and thus the two eigen values $\lambda_{\pm}$ have opposite signs. 
$W_{11}W_{22} - W_{12}^2 > 0$ can be proved as follows. 
First, we consider an average 
\def\eqavedef{(11)}
$$
     \langle \cdots \rangle 
     \equiv 
     \frac{1}{N} \sum_{\vk} 
     \frac{\tanh\frac{\epsilon_{\vk}}{2T}}{2\epsilon_{\vk}}
     (\cdots) . 
     \eqno\eqavedef
     $$
Then, since an inequality 
$\langle (\gamma_1 x - \gamma_2 )^2 \rangle > 0$ 
is satisfied for arbitrary real number $x$, 
we obtain 
$$
     \langle \gamma_1 \gamma_2 \rangle^2 
     - \langle {\gamma_1}^2 \rangle \langle {\gamma_2}^2 \rangle < 0 . 
     $$
Thus, $W_{11}W_{22} - W_{12}^2 > 0$ is proved.

\begin{figure}[htb]
\begin{center}
\leavevmode \epsfxsize=7cm  \epsfbox{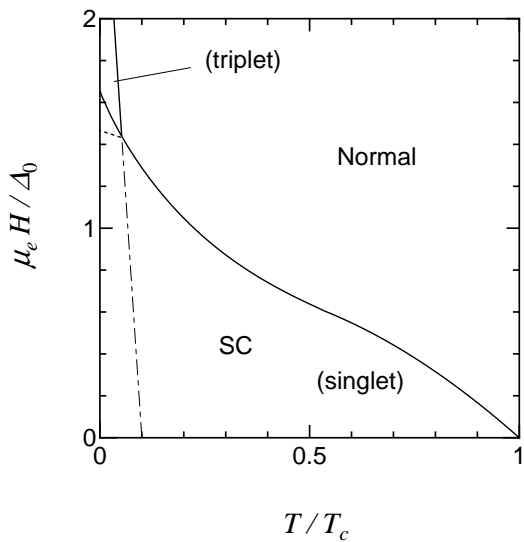}
\end{center}
\caption{
Schematic phase diagram on a temperature and magnetic field 
plane. The vertical solid line at high fields shows the transition 
temperature of a triplet superconductivity of parallel spin pairing. 
The solid curve shows the critical field of singlet superconductivity, 
which is lowered into the dotted line below the critical field 
of the triplet superconductivity at low temperatures. 
The dotted-broken line is the transition temperature of the triplet 
superconductivity when the singlet pairing is suppressed. 
}
\label{fig:phased}
\end{figure}

The fact that the two eigen values have opposite signs means that 
both the singlet and triplet interactions have attractive channels. 
For antiferromagnetic fluctuations, the attractive triplet interaction 
is weaker than the attractive singlet interaction 
as we discussed above. 
Thus, usually, the singlet pairing should occur at zero field. 
However, at high fields where the singlet pairing is suppressed, 
the triplet pairing takes place. 
With an assumption that magnetic field of the order of 
$\mu_e H \sim \Delta_0$ does not change the feature of the spin 
fluctuations drastically, 
we could predict a phase diagram as shown in Fig.\ref{fig:phased} 
on a $T$-$H$ plane.

Since the magnetic field is parallel to the highly conductive layers, 
the orbital pair-breaking effect is not efficient. Thus, the transition 
temperature of the triplet superconductivity of parallel spin pairing 
is reduced only slightly by the magnetic field. 
Hence, the triplet superconductivity appears at high fields.

The upturn of the critical field (${\partial^2 H_c}/{\partial T}^2 > 0$) 
of singlet superconductivity at low temperatures in Fig.\ref{fig:phased} 
is due to an appearance of Fulde-Ferrell-Larkin-Ovchinnikov state 
(FFLO state or LOFF state)~\cite{Ful64,Lar64}. 
In low dimensional systems, the FFLO state is enhanced 
by Fermi-surface effects at low temperatures~\cite{Shi94,Miy99}. 
A reentrant transition to a triplet superconductivity might be observed 
at higher fields~\cite{Leb86}, but the transition temperature could 
recover only up to the zero field value at most.

Our schematic phase diagram is consistent with the experimental results 
in \TMTSFClO~\cite{Lee95} and \TMTSFPF~\cite{Lee97}. 
The observed critical field also shows an upturn at low temperatures 
(${\partial^2 H_c}/{\partial T^2} > 0$), 
and tends to increase divergently in a low temperature limit 
(${\partial H_c}/{\partial T} \rightarrow -\infty$), 
although the observated transitions have widths and are rather ambiguous.

In the FFLO state, the singlet and the triplet order parameters are mixed 
in general, since the finite center-of-mass momentum $\vq$ of the pairs 
breaks a symmetry in the real space, and at the same time the magnetic 
field breaks the rotational symmetry in the spin space. 
Hence, the FFLO state includes a component of a triplet superconductivity 
of antiparallel spin pairing. 
Matsuo {\it et al.} examined the mixing of the order parameters in an FFLO 
state, and found that the critical field is remarkably enhanced~\cite{Mat94}. 
In particular, the tri-critical temperature of the normal, BCS, FFLO phase 
is enhanced.

The lower critical fields of the FFLO state 
and the tri-critical point are not drawn in Fig.\ref{fig:phased}, 
since it is difficult to assert their locations. 
It is also difficult to predict the orders of the 
transitions at the lower critical fields. 
In \TMTSFClO, since first order transitions below the upper critical 
fields have not been observed, 
the FFLO state might be suppressed in this compound by impurity 
scatterings or some other reasons. 
In such a situation, the upturn of the critical field does not occur 
in our phase diagram. 
Such a suppression of the FFLO state might be a reason of the widths 
of the transitions at the upper critical fields in \TMTSFClO.

In our phase diagram, we should be careful about the choice of the ratio 
of the transition temperatures of the singlet pairing and the triplet 
pairing at zero field. 
The transition temperatures are extremely sensitive to the strengths of 
the pairing interactions, Coulomb repulsive interactions, 
impurity scatterings for anisotropic pairings, and so on. 
For example, if the triplet pairing interactions are too weak 
although they exist, 
triplet superconductivity at high fields is not observed in practice.

On the other hand, the result of the recent Knight shift 
measurements in \TMTSFPF~\cite{Lee00} might be explained by taking 
a larger ratio of the transition temperatures. 
In this compound, triplet superconductivity is supported by the Knight 
shift measurements in a strong magnetic field $B = 2.38{[\rm T]}$, 
and the superconducting transition temperature was still about $40 \%$ 
of the zero field transition temperature $\Tc^{(0)} = 1.18{[\rm K]}$ 
in spite of the high magnetic field. 
If the superconductivity at zero field is due to a singlet pairing 
mediated by antiferromagnetic fluctuations, 
the strong magnetic field $B = 2.38{[\rm T]}$ may favor a triplet pairing, 
since the pairing interactions include attractive triplet channels.

We have illustrated the existence of an attractive triplet channel 
in the interactions mediated by antiferromagnetic fluctuations 
in restricted cases. 
However, it is plausible that this feature is general. 
For example, the interactions between distant sites were calculated 
in a perturbation theory based on the quasi-one-dimensional Hubbard model, 
and it was found that triplet pairing interactions can be 
attractive~\cite{Shi89}.

In a 1/4-filled Hubbard model with electron dispersion 
\def\eqepsdef{(12)}
$$
     \epsilon_{\vk} = - 2 t \cos k_x a - 2 t' \cos k_y b - \mu , 
     \eqno\eqepsdef
     $$
a perturbation theory within an RPA gives pairing interactions 
$V(\vR_{ij} = m {\va} + n {\vb}) \equiv V_{mn}$ 
shown in Table \ref{table:1} and \ref{table:2}, 
where $\va$ and $\vb$ are lattice vectors. 
In these tables, we find oscillations of the interactions in the 
real space. 
For triplet pairing, for example, $V_{10} < 0$ and $V_{40} < 0$ are 
attractive interactions. 
The former favors the intra-chain nearest-neighbor component 
of the gap function $\Delta_{10} \sin(k_x a)$, 
which does not have a node on the open Fermi-surfaces. 
On the other hand, the latter favors the gap function of the form 
$\Delta_{40} \sin(4 k_x a)$, which has line nodes on the Fermi-surface 
near $\vk = (\pm \pi/4a, \pm \pi/2b)$ for 1/4-filling. 
In practice, a superposition up to more distant components 
gives the gap function localized around the Fermi-surfaces with a width 
related to the spin fluctuations.

\begin{table}
\caption{
Singlet pairing interactions $V_{nm}/t$ 
in a quasi-one-dimensional Hubbard model with $t' = 0.150 t$ 
and $U \approx 0.148t$. 
}
\label{table:1}
\begin{tabular}{r|rrr}
         ~&~   $n=0$   ~&~  $n=1$     ~&~  $n=2$    ~ \\
\hline 
\\[-10pt]
 $m = 0$ ~&~ $ 2.769$  ~&~  $-0.067$  ~&~  $ 0.020$ ~ \\
     $1$ ~&~ $ 0.181$  ~&~  $ 0.006$  ~&~  $-0.002$ ~ \\
     $2$ ~&~ $-0.465$  ~&~  $ 0.087$  ~&~  $-0.031$ ~ \\
     $3$ ~&~ $-0.084$  ~&~  $-0.004$  ~&~  $ 0.004$ ~ \\
     $4$ ~&~ $ 0.252$  ~&~  $-0.102$  ~&~  $ 0.036$ ~ \\
     $5$ ~&~ $ 0.056$  ~&~  $-0.002$  ~&~  $-0.005$ ~ \\
$\cdots$ ~&  \multicolumn{3}{c}{$\cdots\cdots$}
\end{tabular}
\end{table}

\begin{table}
\caption{
Triplet pairing interactions $V_{nm}/t$ 
in a quasi-one-dimensional Hubbard model with $t' = 0.150 t$ 
and $U \approx 0.148t$. 
}
\label{table:2}
\begin{tabular}{r|rrr}
         ~&~   $n=0$    ~&~  $n=1$     ~&~  $n=2$    ~ \\
\hline 
\\[-10pt]
 $m = 0$ ~&~  $ 0.768$  ~&~  $ 0.024$  ~&~  $-0.013$ ~ \\
     $1$ ~&~  $-0.078$  ~&~  $-0.002$  ~&~  $ 0.001$ ~ \\
     $2$ ~&~  $ 0.187$  ~&~  $-0.031$  ~&~  $ 0.010$ ~ \\
     $3$ ~&~  $ 0.032$  ~&~  $ 0.002$  ~&~  $-0.001$ ~ \\
     $4$ ~&~  $-0.100$  ~&~  $ 0.036$  ~&~  $-0.012$ ~ \\
     $5$ ~&~  $-0.020$  ~&~  $ 0.001$  ~&~  $ 0.002$ ~ \\
$\cdots$ ~& \multicolumn{3}{c}{$\cdots\cdots$}
\end{tabular}
\end{table}

In conclusion, we point out that the pairing interactions mediated by 
antiferromagnetic fluctuations necessarily accompany attractive 
channels of triplet pairing interactions. 
Therefore, in quasi-low-dimensional type II superconductors 
in magnetic fields parallel to the highly conductive layers, 
if the origin of the pairing interactions is the exchange of 
antiferromagnetic fluctuations, a triplet pairing occurs at high fields 
where the singlet pairing is suppressed by the Pauli pair-breaking effect. 
We propose a schematic phase diagram which may explain the experimental 
observation in \TMTSFClO.

We have discussed quasi-low-dimensional systems in a strong parallel 
magnetic field. 
However, if some extra mechanism suppresses the singlet pairing 
more than the triplet pairing, 
antiferromagnetic fluctuations can give rise to a triplet pairing 
superconductivity even at zero field, or even in three dimensional systems. 
For example, intersite Coulomb repulsions may suppress the pairing 
between electrons on nearest- and next-nearest-neighbor sites. 
Also, the suppression of anisotropic superconductivity due to impurity 
scatterings would be different depending on the pairing anisotropy. 
The quasi-one-dimensional organic superconductors \TMTSFX 
\hspace{0.25ex} have open Fermi-surfaces. 
Thus, the triplet pairing of the form $\Delta(\vk) \sim \sin (k_x a)$, 
which was discussed by Belin {\it et al.}~\cite{Bel97}, 
does not have a node on the Fermi-surface. 
Such a nodeless gap function might have some advantage in comparison 
to the other anisotropic gap functions with line nodes. 
These possibilities will be examined in a separate paper.

This work was supported by a grant for CREST from JST.


\end{document}